\documentclass[10pt]{article}

\usepackage{fullpage}
\usepackage{url}

\usepackage[colorinlistoftodos]{todonotes}
\usepackage{xspace}
\usepackage{fancyvrb}

\newcommand{\nop}[1]{}

\newcommand{\qirk}{\textsf{QirK}\xspace}

\title{QirK: Question Answering via Intermediate Representation on Knowledge Graphs}

\author{
\begin{tabular}{ccc}
  \hspace{0.25cm} Jan Luca Scheerer$^1$ \hspace{0.25cm}& \hspace{0.25cm} Anton Lykov$^2$ \hspace{0.25cm} & \hspace{0.25cm} Moe Kayali$^2$ \hspace{0.25cm} \\[.25em]
  \hspace{0.25cm} Ilias Fountalis$^3$ \hspace{0.25cm} & \hspace{0.25cm} Dan Olteanu$^4$ \hspace{0.25cm} & \hspace{0.25cm} Nikolaos Vasiloglou$^3$ \hspace{0.25cm}\\[.25em]
  \end{tabular}\\
  \begin{tabular}{c}
    \hspace{0.25cm} Dan Suciu$^2$ \hspace{0.25cm}
  \end{tabular}\\ \\
  $^1$ETH Zurich \hspace{0.2cm} $^2$University of Washington \hspace{0.2cm} $^3$RelationalAI, Inc. \hspace{0.2cm} $^4$University of Zurich
}

\date{Version as of: January 31, 2024}

\begin{document}
\maketitle

\begin{abstract}
    We demonstrate QirK, a system for answering natural language questions on Knowledge Graphs (KG). QirK can answer structurally complex questions that are still beyond the reach of emerging Large Language Models (LLMs). It does so using a unique combination of database technology, LLMs, and semantic search over vector embeddings. The glue for these components is an intermediate representation (IR). The input question is mapped to IR using LLMs, which is then repaired into a valid relational database query with the aid of a semantic search on vector embeddings. This allows a practical synthesis of LLM capabilities and KG reliability.
    
    A short video demonstrating QirK is available at \url{https://youtu.be/6c81BLmOZ0U}.
\end{abstract}

\section{Answering Questions using Knowledge Graphs}

Databases operate under the \emph{Closed World Assumption}.  Their domain is
limited but they can answer complex queries correctly, completely, and
efficiently.  Large Language Models
(LLMs)~\cite{DBLP:journals/corr/abs-2303-08774,DBLP:journals/corr/abs-2302-13971}
support an \emph{Open World Assumption}: they can answer simple,
open-ended queries from an impressively large number of domains.
However, LLMs often return incorrect answers, tend to hallucinate, and
fail to answer complex questions that are easily supported by
databases.  A recent systematic study of their
coverage~\cite{DBLP:journals/corr/abs-2308-10168} found that {\em
  ``existing LLMs are still far from being perfect in terms of their
  grasp of factual knowledge, especially for facts of torso-to-tail
  entities.''}

We argue that Knowledge Graphs (KGs) offer the right middle ground for
answering complex questions about a wide range of domains.  Several large, open source KGs have been developed, e.g., Wikidata~\cite{DBLP:journals/cacm/VrandecicK14},
Dbpedia~\cite{DBLP:conf/semweb/AuerBKLCI07},
Yago~\cite{DBLP:conf/www/SuchanekKW07};
see~\cite{DBLP:journals/ftdb/WeikumDRS21} for a recent survey.  Open-source KGs are curated by the community, constantly updated
and improved, and their coverage, while not as large as that of
LLMs, is still impressive.  KGs can naturally be stored in a database and queried with SPARQL or SQL. They can give accurate and complete answers to complex queries in many domains.

\nop{For example, how is a recipient of the Turing award represented?  Is it a
single triple, like $(X, \text{received-award}, \text{Turing Award})$?
The answer is no, even such a simple fact is represented with multiple
links, see Sec.~\ref{sec:examples}}

However, queries over KGs are very difficult to formulate.
Wikidata has over 10,000 distinct properties and well over 100 million entities.  \textit{De facto} these constitute the {\em schema} of the database, and it is impossible, even for an expert, to be familiar with all of them.
\nop{Another significant difficulty is that one needs to be familiar with
how the data is represented, in order to formulate a query. For example, to retrieve Turing Award winners we must lookup the identifier for \texttt{Turing Award} (Q185667) and \emph{know} that such a relationship is described via the \texttt{award received} (P166) property, see Sec.~\ref{sec:examples} for a more detailed example.}  While KGs hold immense potential for a wide array of applications, their extensive and intricate schemas, however, pose a significant challenge as  existing querying systems usually demand an intimate
familiarity with the specific labels and structures in the
graph.

We introduce \qirk ({\bf\underline Q}uestion Answering via {\bf\underline I}ntermediate {\bf\underline R}epre\-sentation on {\bf\underline K}nowledge Graphs), a system that uses a unique combination of LLMs, semantic search on vector embeddings, and database technology to answer complex questions on KGs.  \qirk addresses the previous challenge head-on, by allowing users to formulate both natural language questions and loosely structured, intuitive queries with minimal natural language constraints. It uses an LLM to translate from natural language questions to an intermediate representation (IR), and a FAISS (Facebook AI Similarity Search)~\cite{douze2024faiss} index to resolve keywords in the IR to entities and properties in the KG. By combining the strengths of LLMs and databases, \qirk paves the way for more effective and intuitive KG querying.

Recently, several systems have been developed that
translate natural language to SQL. They often achieve over 90\%
accuracy on Spider~\cite{DBLP:conf/emnlp/YuZYYWLMLYRZR18}, the
standard benchmark for NLP to SQL translation.  However, these systems, e.g., 
CatSQL~\cite{10.14778/3583140.3583165}, are primarily developed for relational databases with a given, and relatively small schema. Newer systems ~\cite{xu2023finetuned} have demonstrated the capability to directly generate SPARQL queries. This poses greater difficulty for LLMs and is less transparent to users and limited by its requirement to link keywords to a single entity/property.
\qirk overcomes this by using a much simpler intermediate representation that can be repaired flexibly into valid SPARQL and SQL queries.




\begin{figure*}
  \includegraphics[width=\textwidth]{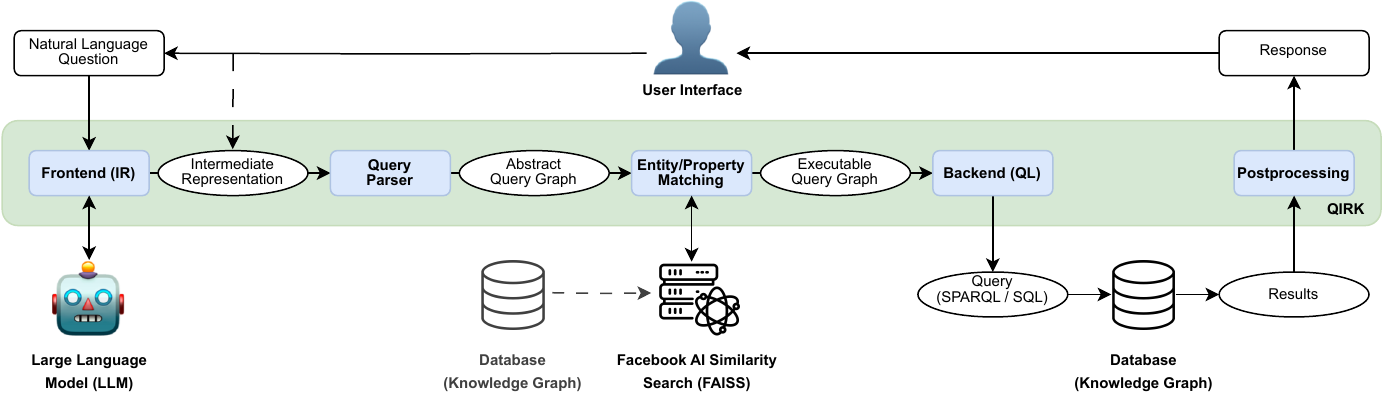}
  \caption{\qirk Architecture. The natural language question is translated into an intermediate representation (IR) with the help of an LLM. Keywords in the IR are  resolved to identifiers of semantically matching KG entities and properties to obtain an executable query graph, which is translated into SPARQL or SQL and evaluated using a database system hosting the KG.}
  \label{fig:architecture}
\end{figure*}

\section{\qirk Architecture}

Fig.\@~\ref{fig:architecture} overviews the architecture of \qirk. 
The user can pose natural language questions in \qirk's user interface. Using an LLM\footnote{The current implementation of \qirk uses ChatGPT 3.5~\cite{DBLP:journals/corr/abs-2005-14165}.}, e.g., ChatGPT (\url{https://chat.openai.com}), Bard (\url{https://bard.google.com}), or Copilot (\url{https://www.bing.com/chat}), the question is translated into an intermediate representation (IR). The IR is a formal language, but still uses snippets of natural language to referr to  entities, relationships between entities, properties, and qualifiers.
\qirk internally represents the IR as an abstract query graph, where relationships (relational atoms) become edges and entities (query constants and variables) become nodes.

The keywords in the IR are then matched against entities and properties available in the KG using an index storing high-dimensional vector embeddings of entities and properties derived from the KG. \qirk requires that this index is pre-computed from the input KG before answering questions over it. Embeddings are extracted using the SentenceBERT model~\cite{reimers-2019-sentence-bert}. This matching step can yield several possible entities and properties that are in the KG and that are similar to the keywords in the IR. Each such matching comes with a similarity score (cosine similarity), which captures the confidence in the quality of the matching: a score value close to 1 (0) signals a good (poor) matching.

Using these matches, \qirk assembles an executable query graph specialized to the vocabulary of the underlying KG. This graph is translated into a structured query language supported by the underlying database system: we currently translate to SPARQL, and, from there, to SQL (as our KG is stored in PostgreSQL). Both  SPARQL and SQL queries use correct identifiers for entities and properties in the input KG. They are not human readable, which makes a direct expression of such queries challenging for humans. 

In the last step, the query is evaluated using a database system that hosts a relational encoding of the KG. The current implementation of \qirk uses PostgreSQL. The query answers are presented to the user in a tabular form with additional confidence information. This confidence is the average score of matches of the keywords in the IR to actual entities and properties in the KG. As there may be several possible matches for the same keyword, each with its own score, there can be answers that are produced by different matches and therefore have different overall scores. \qirk ranks the answers by inspecting the scores of the matched properties and entities.

\nop{ \qirk picks the query with the largest score, which is the average of the largest scores for the semantic matching of each keyword in the IR to an entity or property.}

Even though the obtained SPARQL/SQL query contains a list of potential entity/property identifiers for each keyword in the IR, it nevertheless remains selective. This is because most combinations of entities and properties from the lists associated with the different keywords in the query are not logically coherent and, thus, do not produce valid relationships in the KG. Therefore, \qirk implicitly leverages the KG to identify semantically coherent combinations without requiring prior knowledge of meaningful associations.

\nop{\qirk does not know in advance which of these combinations make sense, it simply relies on the relationships between entities and properties existing in the KG to resolve this semantic matching.}

\section{\qirk by Examples}
\label{sec:examples}

Let us first consider the following question: 

\begin{quote}
{\em "Name people who have won both an Oscar for Merit and a Turing Award."}
\end{quote}

 Current LLMs fail to answer this query (as of January 27, 2024): ChatGPT is unaware of any such individuals. Bard returns incorrect results (e.g., \texttt{Steven Spielberg}), while Copilot states it cannot find any such people. In contrast, \qirk uses the Wikidata KG to provide the correct answer: \texttt{Edwin Catmull}.

\qirk first translates the natural language question into an IR:
\begin{Verbatim}[frame=single, commandchars=+\[\]]
    X: received_award(X, "Oscar for Merit"); 
       received_award(X, "Turing Award")
\end{Verbatim}
The IR syntax resembles first-order logic: $X$ is a variable and we seek values for $X$ such that the body of the query is satisfied, i.e., the relationship \texttt{received\_award} holds between $X$ and both \texttt{"Oscar for Merit"} and \texttt{"Turing Award"}. 

Evaluating a SPARQL or SQL encoding of this IR on Wikidata yields no answer, since  neither the property \texttt{received\_award} nor the entity \texttt{Oscar for Merit} exist in the KG. \qirk addresses this by using the FAISS index to resolve predicates and entities in the KG that are semantically similar to those in the IR. To improve the accuracy of the matching, \qirk takes the label, description, and the popularity into account. 
In this example, \qirk resolves the predicate \texttt{received\_award} to the intended property \texttt{award received} (P166).
It also finds the entities \texttt{Medal of Merit} (Q7408872, score=$0.70$), \texttt{Medal for Merit to Culture} (Q1702885, score=$0.74$), \texttt{Medal for Merit} (Q1307005, score=$0.68$), and \texttt{Gold Medal of Merit in the Fine Arts} (Q3753203, score=$0.73$) as well as the intended entity \texttt{Academy Award of Merit} (Q8624, score=$0.77$) as potential candidates for the keyword \texttt{"Oscar for Merit"}. 
For the keyword \texttt{"Turing Award"}, \qirk retrieves the matching entities: \texttt{Turing machine} (Q163310, score=$0.65$), \texttt{Category:Turing Award} (Q9241105, score = $0.67$), \texttt{Turing} (Q490481, score = $0.67$), \texttt{Alan Turing} (Q7251, score=$0.70$), and the intended entity \texttt{Turing Award} (Q185667, score=$0.89$).

\qirk rewrites the IR into an executable query graph, where each keyword is replaced by the list of the retrieved identifiers. It then compiles the query graph into SPARQL or SQL and evaluates it using a relational database system. We next show the generated SPARQL query (the SQL query is similar, albeit more verbose):
\begin{Verbatim}[frame=single, commandchars=+\[\]]
+textbf[SELECT] ?A1 ?C3 ?A2 ?H0 +textbf[WHERE] {
  ?H0 ?C3 ?A1. ?H0 ?C3 ?A2.
  +textbf[FILTER] (?A1 +textbf[IN] (wd:Q7408872, wd:Q8624, wd:Q1702885, 
                  wd:Q1307005, wd:Q3753203) 
       && ?C3 +textbf[IN] (wdt:P166) 
       && ?A2 +textbf[IN] (wd:Q163310, wd:Q185667, wd:Q490481, 
                  wd:Q9241105, wd:Q7251)) }
\end{Verbatim}

The above SPARQL query has the following variables: \texttt{?H0} iterates over individuals, \texttt{?A1} and \texttt{?A2} iterate over the possible entities matched to \texttt{Oscar for Merit} and \texttt{Turing Award} respectively, and \texttt{?C3} is bound to the property \texttt{award received} with identifier P166.

Even though the generated SPARQL query specifies a list of potential matches for each of the two entities \texttt{?A1} and \texttt{?A2}, the conjunction of their relationships with the \texttt{award received} property is only satisfied by logically coherent combinations, namely \texttt{?A1 =  wd:Q8624} and \texttt{?A2 = wd:Q185667}. 
Without \qirk one would have to write the following SPARQL query:

\nop{
To answer the natural language question without using \qirk, we would need to write the following SPARQL query:}

\begin{Verbatim}[frame=single, commandchars=+\[\]]
+textbf[SELECT] ?H0 +textbf[WHERE] {
    ?H0 wdt:P166 wd:Q8624. ?H0 wdt:P166 wd:Q185667.}
\end{Verbatim}

Evidently, this query is significantly harder to express than the natural language question given as input to \qirk.

After the query is evaluated by a database system and the answers are retrieved by \qirk, \qirk re-ranks the answers based on their similarity scores. This is done by analyzing the concrete assignment to \texttt{?A1},  \texttt{?C3} and \texttt{?A2} for each answer and averaging their similarity scores. These results are then presented to the user.

\paragraph{Structurally complex questions.} \qirk can also answer structurally more complex questions, such as: 

\begin{quote}
    {\em "List movies where the director is married to a member of the cast."}
\end{quote}

\begin{figure*}
\centering
\setlength\fboxsep{0pt}
\setlength\fboxrule{0.25pt}
\fbox{\includegraphics[width=\textwidth]{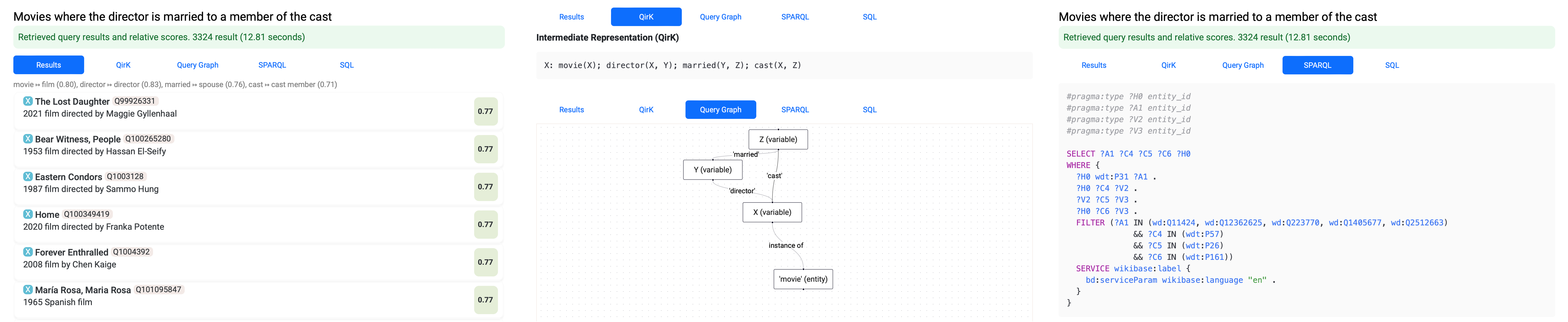}}
\caption{Snapshot of \qirk's user interface. (left) Users can write questions in natural language or in \qirk's intermediate representation. \qirk's answer is presented in tabular form. (middle) The query graph inferred by \qirk from the natural language question is shown for inspection. (right) \qirk displays the SPARQL and SQL queries generated from the executable query graph. The queries include identifiers of entities/properties from the underlying KG obtained using the FAISS index.}
\label{fig:qirk-ui}
\end{figure*}

Figure~\ref{fig:qirk-ui} (left) shows an excerpt of \qirk's answers to this question. \qirk finds 3324 movies that satisfy the intricate triangle relationships between movies, directors, and their spouse actors. 

Large Language Models typically provide a small set of answers to such questions. For instance, ChatGPT only gives four (correct) answers. Bard gives 20 answers, several of them are partially incorrect as they do not satisfy the three binary relationships. Such answers correctly identify actors and cast members involved in the same movie, yet without being married to each other, e.g.: (i) 1990: Pretty Woman (Director: Garry Marshall, Cast: Julia Roberts); (ii) 1994: Pulp Fiction (Director: Quentin Tarantino, Cast: Uma Thurman); (iii) 2001: The Lord of the Rings: The Fellowship of the Ring (Director: Peter Jackson, Cast: Liv Tyler).
Bing Chat is able to retrieve a website\footnote{\url{https://rb.gy/zyg6lx}} with concrete examples and therefore provides a limited number (ten) of correct results.
A clear difficulty of using LLMs for such questions is their unreliability and the time-consuming effort to check whether their answers are correct. In contrast, \qirk observes all relationships stated in the question and can only provide answers that satisfy all these relationships.


\qirk expresses this question in IR as the following cyclic query:

\begin{Verbatim}[frame=single, commandchars=+\[\]]
X: movie(X); director(X,Y); married(Y,Z); cast(X,Z)  
\end{Verbatim}

Using the vector index, \qirk retrieves the following semantically similar properties (and identifiers) for the keywords in the IR. For the keyword \texttt{movie}, it finds: \texttt{Movie} (Q2512663) (disambiguation page), \texttt{film} (Q11424), \texttt{Film} (Q12362625) (album), \texttt{B movie} (Q223770) (genre), and \texttt{Movie Movie} (Q1405677) (film by Stanley Donen). The keywords \texttt{director}, \texttt{married}, and \texttt{cast} are mapped to the properties \texttt{director} (P57), \texttt{spouse} (P26), and \texttt{cast member} (P161) respectively.
\qirk combines these matches with the IR to generate the SPARQL query (also  depicted in Fig.\@~\ref{fig:qirk-ui}):

\begin{Verbatim}[frame=single, commandchars=+\[\]]
+textbf[SELECT] ?A1 ?C4 ?C5 ?C6 ?H0 +textbf[WHERE] {
?H0 wdt:P31 ?A1. 
?H0 ?C4 ?V2. ?V2 ?C5 ?V3. ?H0 ?C6 ?V3.
+textbf[FILTER] (?A1 +textbf[IN] (wd:Q2512663, wd:Q11424, wd:Q12362625, 
                wd:Q223770, wd:Q1405677) 
     && ?C4 +textbf[IN] (wdt:P57) 
     && ?C5 +textbf[IN] (wdt:P26) 
     && ?C6 +textbf[IN] (wdt:P161)) }    
\end{Verbatim}

\paragraph{Datatypes.} With the exception of globe coordinates, \qirk supports all Wikidata datatypes: \texttt{entity\_id}, \texttt{string}, \texttt{date}, \texttt{numeric}, and \texttt{quali\-fiers}. \qirk resolves the datatypes automatically, where possible, or uses those supplied in the IR. The type of a query variable must be consistent across the query, but need only be declared for one of its occurrences.
For instance, the question: "When was Alan Turing born?" is transformed into the following IR:
\begin{Verbatim}[frame=single, commandchars=+\[\]]
X: date_of_birth("Alan Turing", X / date)
\end{Verbatim}

We can also ask for the US presidents and their heights:
\begin{Verbatim}[frame=single, commandchars=+\[\]]
X, Y: position(X, "President of the United States");
      height(X, Y / numeric)
\end{Verbatim}

where we explicitly state that the variable \texttt{Y} is of type \texttt{numeric}. 

\paragraph{Qualifiers.} \qirk supports qualifiers in its intermediate representation.
For instance, the question "When did Barack Obama become president?" can be expressed as follows in \qirk's IR:

\begin{Verbatim}[frame=single, commandchars=+\[\]]
Y: X := holds_position("Barack Obama", "President");
   start_time(X, Y / date)
\end{Verbatim}

The qualifier $X$ is assigned to the fact \texttt{holds\_position("Barack Obama", "President")} and used to make a further statement.
This statement is expressed by the second relational atom in the IR, where we ask for the \texttt{start\_time} of the previous fact.

\paragraph{Aggregation.} \qirk also supports min/max aggregation. For instance, we can ask for the height of the tallest US president:

\begin{Verbatim}[frame=single, commandchars=+\[\]]
MAX(Y): position(X,"President of the United States"); 
        height(X,Y / numeric)
\end{Verbatim}

\qirk's functionality can be extended to support further aggregates such as count, sum, and average. \nop{While queries asking for a precise quantity might not make sense in our incomplete information setting, it may still be useful to know whether a quantity asked as a question is above a given threshold.}

\section{User Interaction}

\qirk provides a web user interface. Fig.\@~\ref{fig:qirk-ui} depicts three snapshots of \qirk answering the movie question from Sec.\@~\ref{sec:examples}. \qirk's interface has one tab to depict each transformation step of the user question, from the initial statement in natural language to the intermediate representation, and finally to SPARQL and SQL queries.

On the {\em Search} tab (not shown), the user can enter the question either in natural language or in \qirk's intermediate representation. 

On the {\em Results} tab, the user can inspect the answers. \qirk groups the answers by the mapping from the keywords in the question to KG entities and properties. All answers in a group have the same confidence score. Both the mapping and the associated score are presented for each answer group. Each answer comes with an entity identifier and a link to the Wikidata page describing it.

The {\em Intermediate Representation} tab graphically depicts the IR query graph made up of the query variables and relations. \nop{Whenever available, the variables are annotated with type information (e.g., integer, date, string). }

The {\em SPARQL} and {\em SQL} tabs show the final structured queries. Fig.\@~\ref{fig:qirk-ui} (right) shows the SPARQL query for the movie question. \nop{The SQL queries are more verbose as they need to explicitly express the joins, which are implicit in SPARQL and the IR.}

\section{Demonstration Scenarios}

The demonstration will show how \qirk answers a combination of ad-hoc and previously prepared questions designed to highlight specific strengths and weaknesses of \qirk:

{\bf Reliability.}
\qirk can provide high-quality answers to questions where LLMs fail (hallucinate or are not capable of providing an answer). Examples of such questions include asking for people or facts that satisfy a conjunction or disjunction of properties, such as asking for people who won as Oscar and a Turing award (Sec.\@~\ref{sec:examples}).

{\bf Interpretability.}
\qirk allows users to scrutinize each stage of the retrieval process. Through the transparent presentation of the explicit mapping of keywords to entities/properties in the KG, users can assess the accuracy of results. This stands in stark contrast to LLMs, which offer no visibility into the retrieval stages.

{\bf Robustness.}
\qirk is robust to minor changes to the question that preserve its meaning such as  replacing keywords with synonyms and accommodating typos, e.g.,: variations such as \texttt{holds\_posi\-tion} instead of \texttt{position}, \nop{\texttt{directed\_by} instead of \texttt{director},} \texttt{actor} instead of \texttt{cast}, or \texttt{Oscar for Merit} instead of \texttt{Academy Award for Merit}. 

{\bf Expressiveness.}
Using aggregates, types, and qualifiers, \qirk can provide answers to complex queries, as highlighted in Sec.\@~\ref{sec:examples}.
    
{\bf Limitations.}
\qirk may not yet offer satisfactory answers to questions that are ambiguous or require extensive reasoning, e.g., "Does it make sense to bring Euro on a trip to the US?".

\nop{ or "Who was the tallest US president married to?"}

\bibliographystyle{abbrv}
\bibliography{main}

\end{document}